\documentstyle[preprint,prb,aps]{revtex}
\begin{document}
\draft
\title{Rotationally symmetric ordered phase in the
\\three-state antiferromagnetic Potts model}

\author{Ralf K. Heilmann$^1$, Jian-Sheng Wang$^2$, Robert H. Swendsen$^1$}
\address{$^1$Department of Physics, Carnegie Mellon University, Pittsburgh,
PA 15213, U.S.A.}
\address{$^2$Computational Science, National University of Singapore,
Lower Kent Ridge Road,
\\Singapore 119260, Republic of Singapore}
\date{\today}
\maketitle
\begin{abstract}

We investigate the three-state antiferromagnetic Potts model on a simple cubic
lattice with a cluster flipping Monte Carlo simulation algorithm in the
temperature region below the transition into disorder at $T_{c1}$.
We find both the well
established broken-sublattice-symmetry (BSS) phase at low temperature and a
new, rotationally symmetric phase at higher temperature, but below $T_{c1}$.
The properties of the second phase and the transition
temperature to the BSS phase are in disagreement with recent cluster-variation
and Monte Carlo simulation results, but in agreement with simulations by
Kolesik and Suzuki.

\end{abstract}


%
%

\section*{I. INTRODUCTION:}

Due to their unusual properties caused by highly degenerate ground states,
antiferromagnetic Potts models have been attracting much attention. Some
recent studies focussed particularly on  the three-state antiferromagnetic
Potts model on a simple cubic lattice. Field theoretical calculations,
renormalization group theory,
Monte Carlo simulations, the cluster-variation method\cite{roslap},
and the coherent-anomaly method\cite{kosu1}
have been
applied to obtain information about its critical behavior. Reference 3
gives a detailed history of developments.

The q-state Potts model is defined by the following Hamiltonian:
$$ H = K \sum_{\langle i,j \rangle} \delta_{s_i, s_j},
\eqno(1) $$
where the spins take on the values $s_i = 1,2,...,q$, the sum is over
nearest neighbor pairs on sites on a lattice, and $\delta$ is the
Kronecker delta.
The lattice can be split up into two sublattices $a$ and $b$, such that all
nearest neighbors of any site belong to the other sublattice.
$K=J/k_B T$ is a dimensionless coupling constant. We set $J/k_B = 1$ and use
$T=1/K$ as a dimensionless reduced temperature.

At low temperatures ($T\ll 1$) the model with q=3 on a simple cubic lattice is
known to be in the so-called broken-sublattice-symmetry (BSS) phase:
one lattice is
mostly occupied by a single state, while the spin values on the other
sublattice are split equally between the remaining two states. The two
sublattices are therefore not equivalent.
Permutation of the three Potts states and the
two sublattices leads to a six-fold degeneracy at low temperatures.

At a temperature of $T_{c1} = 1.22605(5)$ as determined by earlier Monte Carlo
studies \cite{gothas} the model undergoes an order-disorder phase transition,
belonging to the XY model universality class.\cite{prb,gothas,prl}

Not much attention had been given to the medium temperature region below
$T_{c1}$,
until Rosengren and Lapinskas (RL) \cite{roslap},
based on their recent cluster-variation
calculations, claimed the existence of three ordered states of different
symmetry: the BSS phase below $T_{c3}=0.7715$,
a 12-fold degenerate phase between
$T_{c3}$ and $T_{c2}=0.787$, and a permutationally symmetric sublattice (PSS)
phase from $T_{c2}$ to $T_{c1}$.

Spin configurations in the different phases can best be described through
the set of concentrations of Potts
states on the two sublattices $a$ and $b$,
$$c_i^j = {2 \over L^3}\sum_{k \in j}\delta_{s_k,i}, \eqno(2)$$
where $i=1,2,3, j=a,b$, and $L$ is the linear system size.
At very low temperature a typical BSS state would therefore have $c_1^a
\approx 1,
c_2^a\approx 0\approx c_3^a,
c_1^b\approx 0, c_2^b\approx {1\over 2}\approx c_3^b$.
A typical PSS state is postulated to have
$c_1^a=c_1^b > {1\over 3}, c_2^a = c_1^b < c_3^a = c_3^b < {1\over 3}$.

Subsequently, Kundrotas, Lapinskas, and Rosengren (KLR) \cite{klr}
performed standard
Metropolis Monte Carlo simulations, focusing on a sublattice
magnetization
\\ $M_j = {1\over 2}(|p_1^j-p_2^j| + |p_2^j-p_3^j| + |p_3^j-p_1^j|)$,
where j again specifies the two sublattices $a$ and $b$.
If the two sublattices have an
equivalent set of concentrations $c_i^j$ as in the PSS state, then the
difference between the sublattice magnetizations $M = M_a - M_b$
should be equal
to zero, while for a completely ordered BSS state it should be one. For this
reason KLR use $M$ as an order parameter to distinguish between the BSS and the
PSS phase.

They then use a method developed by Lee and Kosterlitz \cite{leekos}
to obtain the
difference in free energy $\Delta F$ between
states of $M=0$ and values of $\pm M$ corresponding to BSS states as a function
of temperature and system size, and conclude from their data that there is a
transition from BSS to PSS at a temperature of $T=0.68(1)$.

We follow a different approach, using the previously defined \cite{prb}
order parameters
$$ \xi_1  = {\sqrt3 \over2} \bigl( c_3^b - c_3^a\bigr),\eqno(3a)$$
$$\xi_2  = {1\over2} \bigl( c_1^a - c_1^b - c_2^a + c_2^b \bigr).\eqno(3b)$$
In a $\xi_1$-$\xi_2$-plane perfect BSS states would lie on a circle around the
origin
at angles of
$\phi = tan^{-1}(\xi_2/\xi_1) = n\pi /3$, while PSS states are at
$\phi = (2n+1)\pi /6$ with $n = 0,\pm 1,\pm 2,...$.

The Fourier component of the angular distribution,
$$\phi_6=\cos\bigl( 6 \tan^{-1}(\xi_2/\xi_1)\bigr),
\eqno(4)$$
should then be $+1$ in the BSS phase and $-1$ in the PSS phase.

\section*{II. RESULTS}

We performed Monte Carlo simulations on simple cubic lattices with linear
sizes of L=8, 16, 32, and 64 and periodic boundary conditions, using
efficient modified cluster-flipping algorithms \cite{prb,prl}.
At least $10^6$ lattice updates or Monte Carlo steps (MCS) were performed
in equilibrium for each temperature and system size.

Fig.\ \ref{histc}
shows the distribution of sublattice concentrations, accumulated into
a histogram. The locations of the
peaks in these distributions correspond to the expectation values of
concentrations of figure 2 in RL. At low temperatures we can see four
peaks with relative weights 1,2,2,1 as expected for the BSS phase.
If the PSS phase were present at higher temperatures, we should be able
to see three peaks with weights 2,2,2. However, the system never develops into
a three-peak distribution. Instead, above $T\approx 0.95$ it shows only
two peaks and a broad distribution between the two peaks.

This broad distribution can be interpreted as result of a uniform angular
distribution in $\phi$.
As we can see in Fig.\ \ref{phi}, this is indeed the case:
at low temperatures the probability distribution of $\phi$ peaks around values
corresponding
to the BSS state and has minima for PSS states.
At higher temperatures the distribution surprisingly becomes uniformly flat,
making no distinction between BSS and PSS or any intermittent states in that
phase.

Finally we show in Fig.\ \ref{phi6t} a plot of the thermodynamic average
$\langle \phi_6 \rangle$ as a function
of temperature and
system size. For all system sizes $\langle \phi_6 \rangle$ is positive at low
temperatures and
decreases with increasing temperature. It becomes zero clearly below the
transition to the disordered phase,
which demonstrates a symmetry-breaking transition
from the BSS state into a phase that is rotationally symmetric in the
$\xi_1$-$\xi_2$-plane.
The strong finite-size dependence of $\langle \phi_6 \rangle $ is very unusual.
This makes an accurate estimate of the transition temperature difficult.
However, we find $\langle \phi_6 \rangle$ to be effectively zero at $T=1.0$
for all system  sizes investigated.
We therefore estimate the transition temperature to lie close to 1.0 in
disagreement with KLR. Our results do agree with those obtained independently
by Kolesik and Suzuki \cite{kolsuz}.

\section*{III. DISCUSSION:}

Our results indicate the following:  there is a transition from the
BSS state into an intermediate state at a temperature $T_{rot}$ lower
than the critical temperature $T_{c1}$.  In the intermediate
region, the distribution of the vector order parameter $(\xi_1, \xi_2)$
is fully
rotational symmetric, very much like the planar rotation or XY models.
There is no region where the
order parameter corresponds to the PSS phase. This actually agrees with the
results shown in figure 2 of KLR, where $|M|$,
which should be zero for a PSS
state, is significantly different from zero for all temperatures below
$T_{c1}$. Fluctuations that would cause $\langle |M| \rangle$ to differ
slightly from zero would also be expected to decrease with
increasing system sizes. This is not the case in figure 2 of KLR.

KLR's histograms of $M$ also show that between $T=1.0$ and $T_{c1}$ there is
always a continuous
range of values of $M$ almost equally accessible to the system,
extending all the way from BSS to
PSS states. Therefore we do not think that their order parameter $M$ is suited
to distinguish between the two ordered phases and locate the transition
temperature.

Qualitatively our results could be explained in the following way:
If one looks at the PSS state as a configuration with, i.e., state 1
dominating sublattice $a$ and state 2 dominating sublattice $b$, one can
start to ``rain down'' randomly state 3 spins onto the whole lattice.
This allows for a higher entropy than the BSS state (where one sublattice
is basically ``frozen out'').   The randomness is restricted however,
since two state 3 spins do not want to be next to each other.  The
increase in entropy comes therefore at a cost in energy.  Our numerical
results indicate, that this trade-off never favors the PSS over the
BSS states, while RL reached a different conclusion at this
point.

\section*{IV. CONCLUSIONS}

In summary we have found evidence for a rotationally symmetric ordered phase in
the three-state AF Potts model on the simple cubic lattice in an intermediate
temperature range between the low temperature BSS phase and the disordered
phase above $T_{c1}$. We estimate the transition between the ordered
phases to take place around $T_{rot}\approx 1.0$, significantly
higher than previously reported.
The exact
location and the nature of this transition need more investigation.

We thank E. Domany for stimulating discussions.


\begin{figure}
\caption{
Summary histogram of the sublattice concentrations $c_i^j$ for linear system
size $L=64$ after $2.5\times 10^6$ MCS per temperature.
The curves for different temperatures have been shifted for clarity.}
\label{histc}
\end{figure}

\begin{figure}
\caption{Histogram of the angle $\phi$ of the vector order parameter
from eqns. (3) for $L=16$ after $1.4\times 10^7$ MCS per temperature.
The curves for different temperatures have been shifted for clarity. }
\label{phi}
\end{figure}

\begin{figure}
\caption{Average of the Fourier component $\phi_6$ as a function of reduced
temperature. Filled circles are for $L=64$,
white squares for $L=32$,
open circles for $L=16$, and triangles for $L=8$. Data points are obtained from
averaging over $2.5\times 10^6, 10^6, 1.4\times 10^7$,
and $3.5\times 10^7$ MCS per temperature respectively.
}
\label{phi6t}
\end{figure}

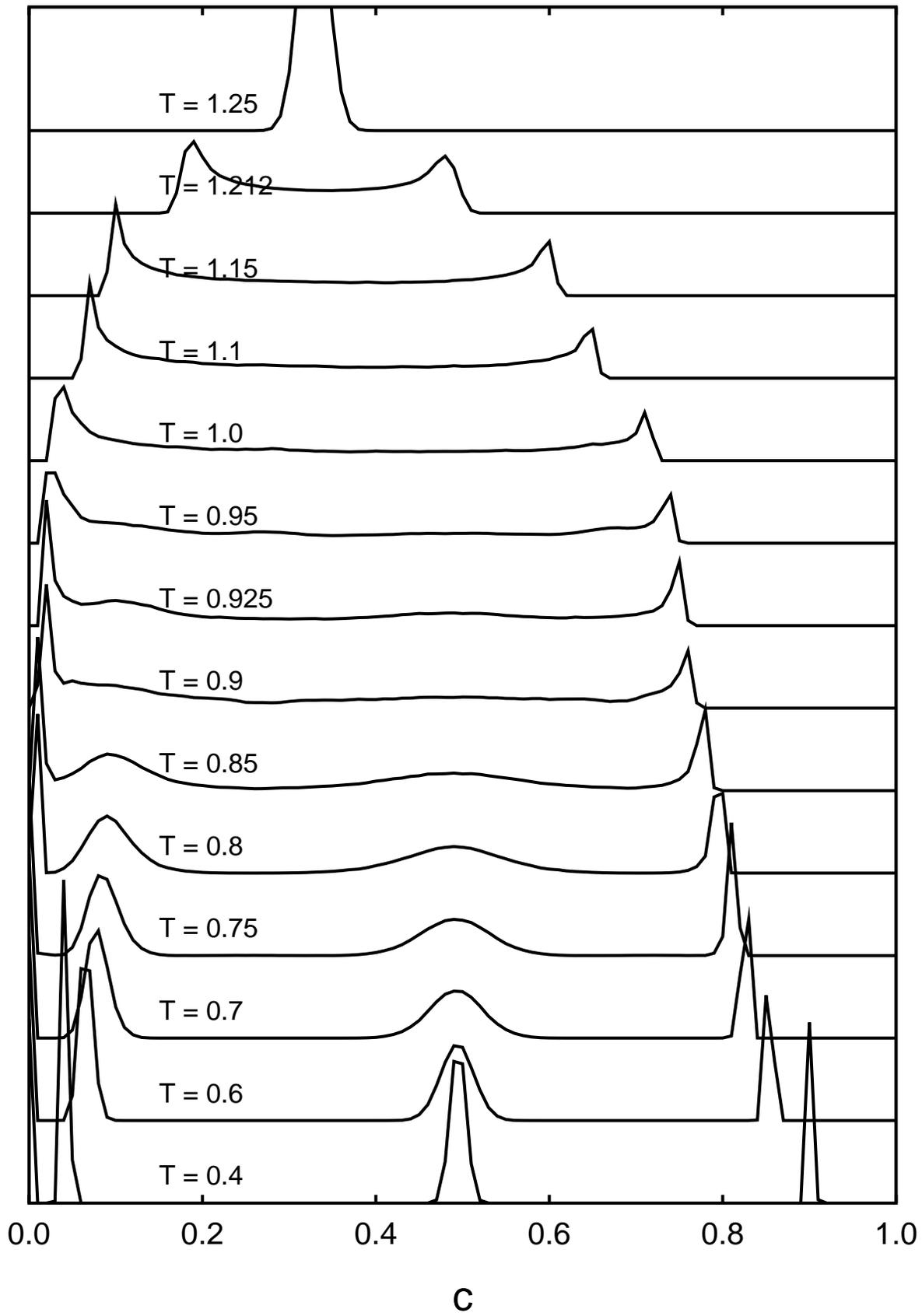
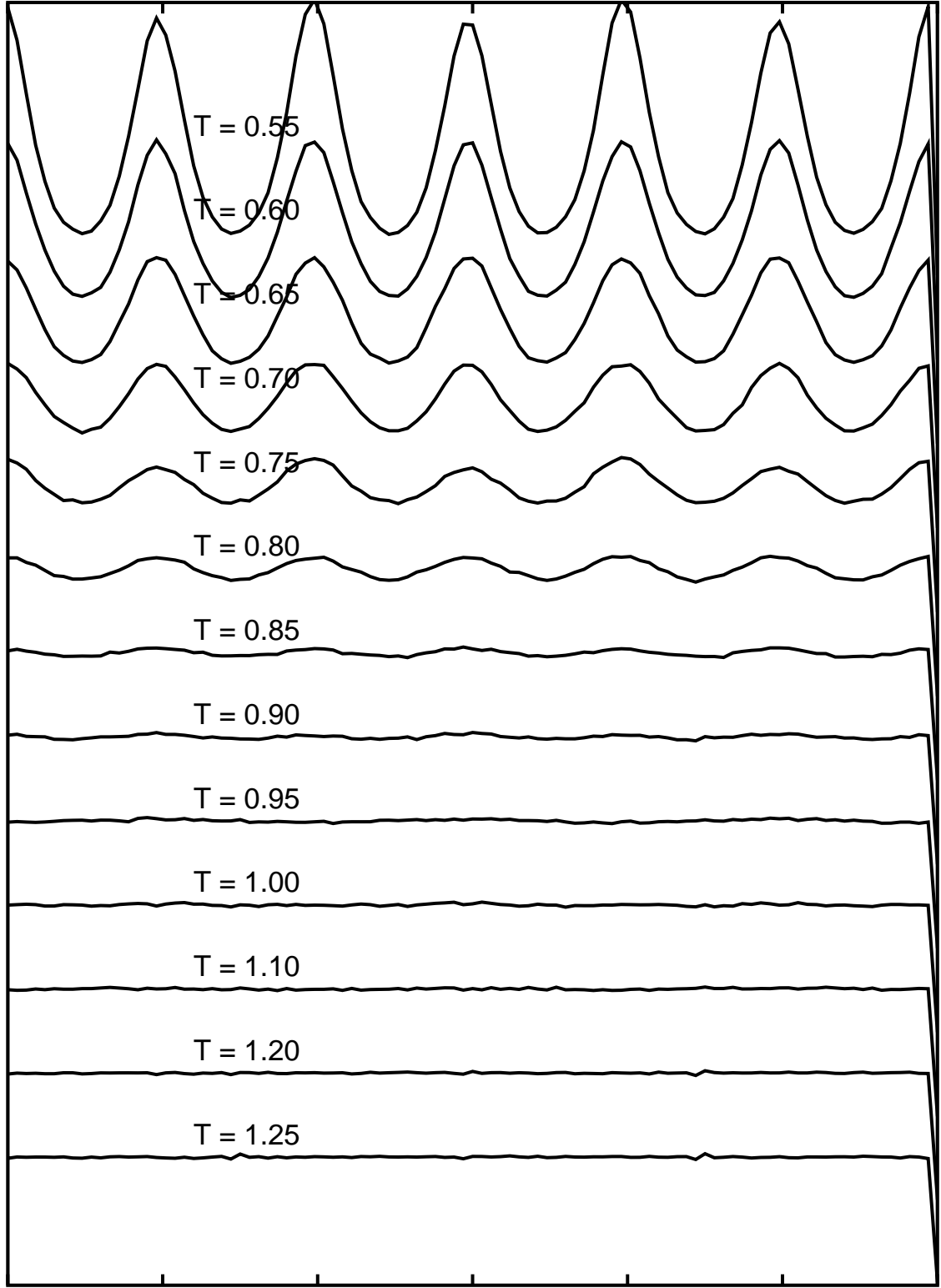
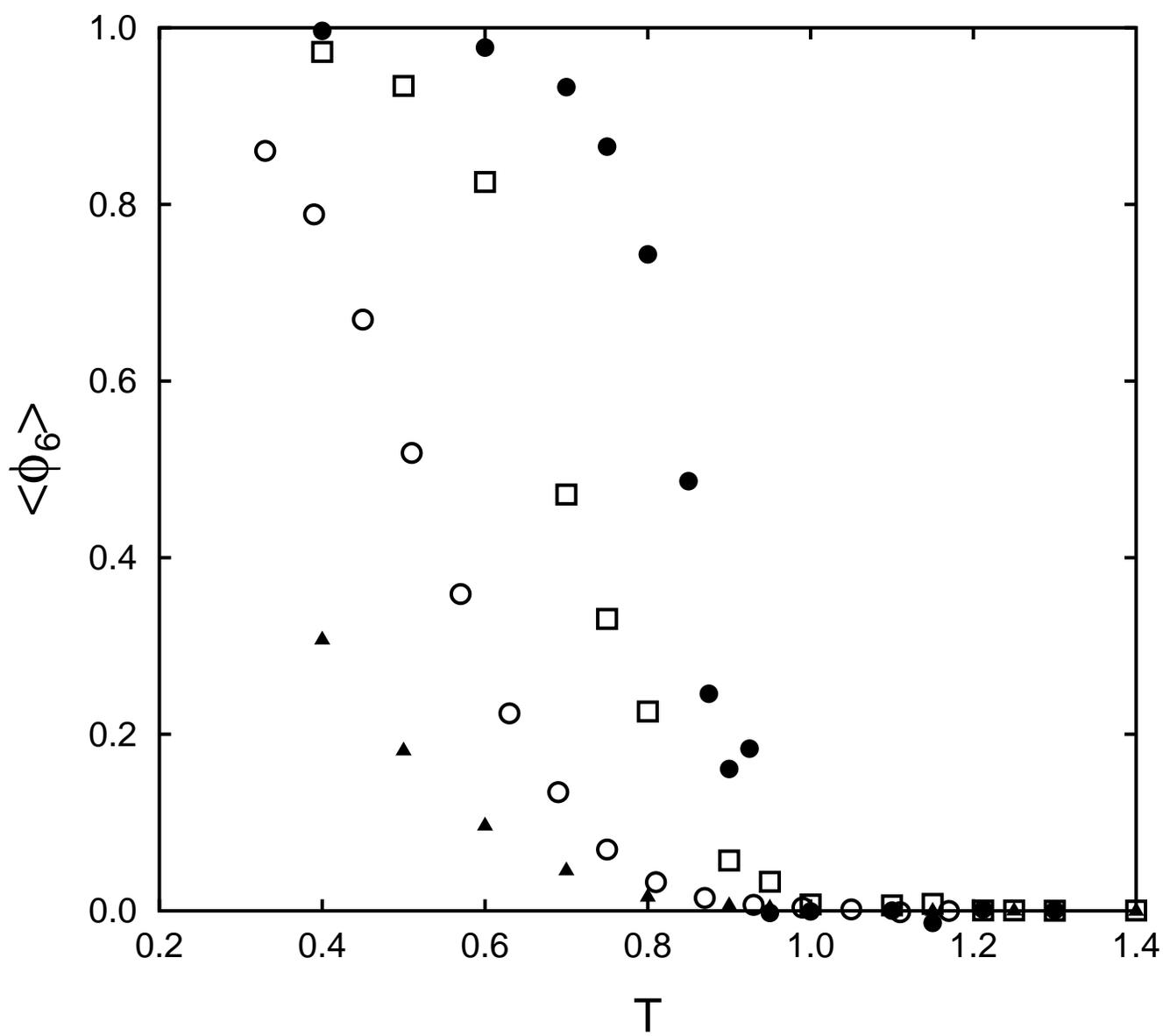
\end{document}